\documentstyle[referee]{l-aa}
\input{epsf}
\input{rotate}
\begin{document}

\thesaurus{04(02.07.1; 11.11.1; 03.13.1; 12.04.1)}
\title{An exact
analytical method for inferring the law of gravity from the
macroscopic dynamics:
Spherical mass distribution with exponential density}
\author{C. Rodrigo-Blanco}
\institute{Laboratorio de Astrof\'{\i}sica Espacial y F\'{\i}sica 
Fundamental (LAEFF),
P.O. Box 50727, E-28080 Madrid, Spain.\\
e-mail: crb@laeff.esa.es}
\date{Received date; accepted date}

\maketitle
\markboth{C.Rodrigo-Blanco: Non-Newtonian gravity directly from 
observations: spherical mass distribution.}{}

\begin{abstract}
We consider the gravitational potential and the
gravitational rotation field generated by an
 spherical mass distribution with exponential density, when the 
force between any two
mass elements is not the usual Newtonian one, but some general 
central force. 
We invert the usual integral relations and
obtain the elemental interaction (between two point-like masses) as
a function of the macroscopic gravitational field (the one generated
 by the distribution).
Thus, we have a direct way for testing the possibility of finding
a correction to the Newtonian law of gravity that can explain the
observed dynamics at large scales without the need of dark matter.
We show that this method can be used even in the case of spiral 
galaxies
with a good level of confidence.

\end{abstract}


\section{\bf Introduction}
\label{sec:introduction}

The dynamical analysis of rotation curves of galaxies, binary galaxies,
clusters of galaxies and the structures known at large scales show
large discrepancies between the observed behaviour and the one expected
from the application of General Relativity and its Newtonian limit to
the visible mass. This disagreement has led many astrophysicists to
believe in the existence of a large amount of non-visible matter and
is, thus, commonly known as the {\it Dark matter problem}.

In spite of this, there is no direct evidence for the validity of
either Einstein's General Relativity or Newtonian gravity at scales 
much larger than those of the Solar system (See, e.g., Will \cite{Will93}).
 There is therefore no experimental
or observational reason to ascertain that unmodified General
Relativity holds at larger distances. This leads us to
 think that we should be open to the possibility  that it had to be
revised (perhaps in the same spirit as Newton's law had to be modified
for strong fields and large velocities).

In this paper we consider the possibility that Newton's law of gravity
is just a good approximation at short distances of a more general
expression for the force. It is interesting to identify which, if any,
extensions of the usual inverse square law are compatible with the
dynamics observed at large scales. 

Work along these lines has already been done (Tolhine 
\cite{Tohline83}, Kuhn
\& Kruglyak \cite{KK87}, Mannheim \& Kazanas \cite{Mannheim89})
assuming a specific functional form for the force, and then evaluating
the field generated by a mass distribution (for instance, a galaxy)
by performing the corresponding three-dimensional integrals.
We present and work out a method that allows us to 
follow the inverse methodology, that is, to infer, 
directly from observations,
 the phenomenological law of
gravity that is able to generate a given macroscopic
gravitational field. We do it for the case of a spherical mass
distribution with exponential density.

In Section \ref{sec:general} we 
give the general definitions that will be used later in Section 
\ref{sec:spherical}
for the case of spherical symmetry with exponential density.
In Section \ref{sec:development} we
show the mathematical basis underlying the results presented in 
Section \ref{sec:spherical}. Finally we offer some conclusions. In 
Appendix \ref{app:diskvsesf} we consider the possibility of 
applying these results as an approximation to the study of spiral 
galaxies  overcoming the fact that they do not show spherical
symmetry. In Appendix \ref{app:math} we list some of the mathematical
identities used in Section \ref{sec:development}.


\section{General definitions}
\label{sec:general}
Let us assume that the gravitational
 potential ge\-ne\-ra\-ted by a point-like mass
does not correspond to the usual Newtonian form  but can be written
in terms of a function $g(r)$ that describes the deviation
 from the Newtonian law, that is,

  \begin{equation}
	\phi (r) \equiv - \frac{G_0 m_1 m_2}{r}g(r).
  \label{punctpot}
  \end{equation}
where $\phi(r)$ is the gravitational potential experienced by two 
point-like particles
of masses $m_1$ and $m_2$ separated by a distance $r$ and $G_0$ is 
the Newton's
constant.

This modification could, for example,
be due to the many body nature of the mass distribution making up 
the galaxy,
quantum corrections, a relativistic theory different from General 
Relativity...
This is irrelevant in what follows.

The force per unit mass is, by definition, the gradient of
 the potential,

  \begin{equation}
	\vec{F}(r) \equiv - \frac{G_0 m_1 m_2}{r^2}g_{\rm eff}(r) 
\frac{\vec{r}}{r},
  \label{punctforce}
  \end{equation}
where we have introduced

  \begin{equation}
	g_{\rm eff}(r) \equiv g(r) - r g'(r).
  \label{ggef}
  \end{equation}

In this way, to find the total potential or the total force
generated by a mass distribution $\Omega$ with density
$\rho(\vec{r})$, one must integrate over the volume spanned by 
$\Omega$ to get:

  \begin{equation}
	\Phi(\vec{R}) = - G_0 \int \int \int_{\Omega} d^3\vec{r} \  \ 
	\frac{g(|\vec{R}-\vec{r}|)}{|\vec{R}-\vec{r}|} \ \  
\rho(\vec{r})
  \label{potomeg}
  \end{equation}
for the potential experienced by a point mass at a distance $R$ 
from the centre of
$\Omega$, and
  \begin{equation}
	\vec{F}(\vec{R}) = - G_0 \int \int \int_{\Omega} d^3\vec{r} 
\ \  
	\frac{g_{\rm eff}(|\vec{R}-\vec{r}|)}{|\vec{R}-\vec{r}|^2} 
\ \  
	\frac{\vec{R}-\vec{r}}{|\vec{R}-\vec{r}|} \ \  \rho(\vec{r})
  \label{forceomeg}
  \end{equation}
for the force.

In the case that the gravitational potential is only a function of the
distance to the centre of the distribution,
it is convenient to introduce two new functions $\Psi(R)$ 
and $\Psi_{\rm eff}(R)$ such that:

  \begin{equation}
	\Phi(R) \ \equiv \ - \ \frac{G_0 M_{\rm tot}}{R} \ \Psi(R),
  \label{potpsi}
  \end{equation}

  \begin{equation}
	\vec{F}(R) \ \equiv \ -\frac{G_0 M_{\rm tot}}{R^2}
	 \ \Psi_{\rm eff}(R) \ \frac{\vec{R}}{R},
  \label{v2psi}
  \end{equation}
and the rotation velocity of a test particle in a circular orbit bound
to the distribution will be:
\begin{equation}
	V^2_{\rm rot}(R) \ \equiv \ \frac{G_0 M_{\rm tot}}{R}
	 \ \Psi_{\rm eff}(R)
  \label{v2psi2}
  \end{equation}
where the auxiliary functions $\Psi_{\rm eff}(R)$ and $\Psi(R)$ 
satisfy 
the following functional relationship:

  \begin{equation}
	\Psi_{\rm eff}(R) = \Psi(R) -R \Psi'(R),
  \label{psiphi}
  \end{equation}

Our goal is to design a procedure where, assuming that $\vec{F}(R)$ 
is known
 (say from observation of the rotation velocity) for all values of 
$R$,
we obtain a $g_{\rm eff}(r)$ that generates the given rotation 
velocity. Or, in other 
words, given the potential as inferred from observations we 
want to find which $g(r)$ 
could have generated it. Actually, what we will find are 
$g(r)$ and $g_{\rm eff}(r)$ as
functions of $\Psi_{\rm eff}(R)$ and $\Psi(R)$ respectively.


\section{Spherical mass distribution with exponential density}
\label{sec:spherical}

In view of the applications that we have in mind, we  study a spherically
 symmetric distribution
with an exponentially decaying density:

\begin{equation}
	\rho(r) \equiv \rho_0 \ e^{-\alpha r}.
\label{density}
\end{equation}

Our ultimate goal is to find
a method to study the discrepancies between the 
observed rotation curves of 
spiral galaxies and the curves predicted by using Newton's 
law of gravity. 
The luminosity
profile of many spiral galaxies can be well fitted assuming 
that the density of 
luminous matter decreases exponentially with distance from the
centre of the galaxy (Kent \cite{Kent87}).
 This is the reason why we are interested in studing
such a density function, even though spiral galaxies are not spherical.

Using Eqs. (\ref{potpsi}), (\ref{v2psi}) and (\ref{density}), 
in Eqs. (\ref{potomeg}) and 
(\ref{forceomeg}), and considering spherical symmetry,
 the two problems sketched in Section \ref{sec:general} can 
be conveniently recast 
in the form of two integral equations:

\vspace{4mm}

\noindent(i) Given $\Psi(R)$, find a function $g(r)$ that 
satisfies the 
 equation:

	\begin{equation}
		\int_0^{\infty}dr\int_0^{\pi}d\theta\int_0^{2\pi}d\phi
		\frac{g(\sqrt{R^2+r^2-2Rrcos\theta})}
		{\sqrt{R^2+r^2-2Rrcos\theta}}
		r^2 sin\theta e^{-\alpha r}
	 = \frac{8 \pi}{\alpha^3}
		\frac{\Psi(R)}{R} \\
	\label{probpot}
	\end{equation}

\noindent and

\vspace{4mm}

\noindent(ii) Given $\Psi_{\rm eff}(R)$, find a function
	 $g_{\rm eff}(r)$ such that:

	\begin{equation}
		\int_0^{\infty}dr\int_0^{\pi}d\theta
		\int_0^{2\pi}d\phi\frac{g_{\rm eff}
		(\sqrt{R^2+r^2-2Rrcos\theta})}
		{(R^2+r^2-2Rrcos\theta)^{\frac{3}{2}}}
		(R-rcos\theta) r^2 sin\theta e^{-\alpha r} 
	= \frac{8 \pi}{\alpha^3}\frac{\Psi_{\rm eff}(R)}{R^2}.
	\label{probforce}
	\end{equation}

In the next section the solution to these integral equations
will be described in detail. The results can be summarized as:

\vspace{4mm}

\noindent(i) Potential problem ({\it viz.} Eqs. (\ref{potpsi}) 
and (\ref{probpot}))

\vspace{4mm}

       In this case, the exact solution to the problem is 
\begin{equation}
		g(x)= 
		\Psi(x)-\frac{2}{\alpha^2}\Psi''(x)+ 
		\frac{1}{\alpha^4}\Psi^{(iv)}(x) 
	\label{gpsi}
	\end{equation}
       where the function $\Psi$ has the following behaviour at 
the origin:

	\begin{equation}
		\Psi(0) \ = \ \Psi''(0) \ = \ 0.
	\label{gpsicond}
	\end{equation}

\noindent(ii)	Force and velocity problem ({\it viz.} Eqs. 
(\ref{v2psi}), (\ref{v2psi2}) and 
	(\ref{probforce})).

\vspace{4mm}

         Here, the exact solution is given by the following expression:

	\begin{equation}
		g_{\rm eff}(x)= 
		 \Psi_{\rm eff}(x)-\frac{2}{\alpha^2}
\Psi_{\rm eff}''(x) 
		+ \frac{1}{\alpha^4}\Psi_{\rm eff}^{(iv)}(x) 
		+ \frac{4}{\alpha^2 x}\Psi_{\rm eff}'(x) - 
		\frac{4}{\alpha^4 x^4}[2x\Psi_{\rm eff}'(x)-
		2x^2\Psi_{\rm eff}''(x)+x^3\Psi_{\rm eff}'''(x)] 
	\label{geffpsieffprob}
	\end{equation}

  The behaviour of $\Psi$ at the origin is as follows:

	\begin{equation}
		\Psi_{\rm eff}(0) \ = \ \Psi_{\rm eff}'(0) \ = \
		 \Psi_{\rm eff}''(0) \ = \ 0.
	\label{geffpsieffcond}
	\end{equation}

The behaviours at the origin just tell us that $\Psi_{\rm eff}(R) 
\propto
R^3$ for $R \sim 0$, and thus, $V_{\rm rot}(R) \propto R$ for $R 
\sim 0$, which is
in fact in good agreement with the observations (as the observed 
rotation curves
are usually well fitted in the inner regions by a straight line) 
and, for a
non-Newtonian gravity point of view, it is also in agreement with the
fundamental experimental constrain that for short
distances the gravitational interaction must be well described by a
Newtonian limit.


\section{Mathematical development}
\label{sec:development}

First, we will show the solution to what we call {\it the potential
problem}, that is, how to go from Eq. (\ref{probpot})
 to Eqs. (\ref{gpsi}) and (\ref{gpsicond}). Later
 we will use these results to solve {\it the
force problem}, i.e, how to go from
Eq. (\ref{probforce}) to Eqs. (\ref{geffpsieffprob})
 and (\ref{geffpsieffcond}).

\subsection{The potential problem.}
\label{subsec:potential}

In order to go from equation Eq. (\ref{probpot}) to Eqs.
 (\ref{gpsi}) and (\ref{gpsicond}), we will need to use several
 mathematical identities. For convenience, these are listed in 
Appendix \ref{app:math}.

Inserting Eqs. (\ref{fourier}), (\ref{addth}), 
(\ref{sinbessel}), (\ref{addth2}) and (\ref{ortog})
 into Eq. (\ref{probpot}) we can write $\Psi(R)$ as:

\begin{equation}
	\Psi(R)= 
	- \frac{\alpha^3}{\pi} \int_0^{\infty}dp
	\frac{\hat{g}_s(p)}{p} \int_0^{\infty} dr e^{-\alpha r} r
	\sin{pr} \sin{pR} 
	= 
	- \frac{\alpha^3}{\pi}
	\frac{d}{d\alpha} \int_0^{\infty}dp
	\frac{\hat{g}_s(p)}{p^2 + \alpha^2}\sin{pR}
\label{inutil1}
\end{equation}

Then, we can apply Eq. (\ref{fourierinv}) and invert the Fourier
transform to obtain, after some straightforward calculations,
 the following more useful form:

\begin{equation}
\begin{array}{rl}
& \\
  \Psi(R)= - \frac{\alpha}{4} & 
	\left\{ (1- \alpha R ) e^{ \alpha R }
	\left[ \int_0^{R}dx  \ g(x) \ e^{-\alpha x} \ - \ 
		\int_0^{\infty}dx  \ g(x) \ e^{-\alpha x}\right] 
	\right.  -   \\
& \\
&	- \ (1+ \alpha R ) e^{ - \alpha R }
	\left[ \int_0^{R}dx  \ g(x) \ e^{\alpha x} \ - \ 
		\int_0^{\infty}dx  \ g(x) \ e^{-\alpha x}\right] +  \\
& \\
&	+ \ \alpha e^{ \alpha R }
	\left[ \int_0^{R}dx  \ g(x) \ x \ e^{-\alpha x} \ - \ 
		\int_0^{\infty}dx  \ g(x) \ x \ e^{-\alpha x} \right] 
	 \ \ +   \\
& \\
&	+ \ \left. \alpha e^{ -\alpha R }
	\left[ \int_0^{R}dx  \ g(x) \ x \ e^{\alpha x} \ + \ 
		\int_0^{\infty}dx  \ g(x) \ x \ e^{-\alpha x}\right]
	    \ \ \ \ \ \right\}   \\
 \end{array}
\label{potexpon}
\end{equation}

In order to simplify these expressions, we introduce an 
au\-xi\-lia\-ry function $\psi(x)$ that makes the integrals exact:

\begin{equation}
	g(x) \ \equiv \ \psi(x) - \frac{2}{\alpha^2} \ \psi''(x)
 	+ \frac{1}{\alpha^4} \ \psi^{iv}(x)
\label{gpsieq}
\end{equation}

We can insert Eq (\ref{gpsieq}) into Eq. (\ref{potexpon}) and, 
upon integration by parts, we get:

\begin{equation}
	\Psi(R) = \psi(R) - (1+\frac{\alpha R}{2}) e^{-\alpha R}
	   \psi(0) + \frac{R}{2 \alpha^2}e^{-\alpha R}\psi''(0)
\label{Psipsi}
\end{equation}
where $\psi$ is a solution to the ordinary differential Eq.
 (\ref{gpsieq}) that satisfies the conditions
 of being
 an analytic function at $x=0$, and

\begin{equation}
	\lim_{x \rightarrow \infty} \ \psi^{(k)}(x) \ x 
 \ \exp(-\alpha x) = 0 \ ;  \ k=0,1,2,3
\label{infcond}
\end{equation}
where  $\psi^{(k)}(x)$ stands for $\psi(x)$ and its three first 
derivatives.

These  conditions are easily fulfilled in all the cases of interest. 
Actually, the analiticity is satisfied in the Newtonian limit, that is
the behaviour that we expect to recover at $R \sim 0$.
Although it is possible to artificially build a {\it pathological} 
$\psi(x)$ such that it can
represent a physical system without satisfying Eq. (\ref{infcond}),
it can be seen that  almost every $\psi$ function that does not 
satisfy it
 corresponds to a  rotation
velocity  that grows almost exponentially with the distance, 
which clearly
seems to contradict the observations.

Actually, it is straightforward to see that, provided 
$\psi(R)$ is a solution to Eq. (\ref{gpsieq}), then $\Psi(R)$ is also a
solution to the same equation. Moreover, the terms proportional to
$\psi(0)$ and $\psi''(0)$ in Eq. (\ref{Psipsi}) assure that $\Psi$ and
its second derivative are both zero at the origin. Taking all 
of this into
consideration, we finally obtain that:

\begin{equation}
	 g(x) =  \Psi(x)-\frac{2}{\alpha^2}\Psi''(x)+
		\frac{1}{\alpha^4}\Psi^{(iv)}(x), 
\label{finalPhi}
\end{equation}
\begin{equation}
	\Psi(0) \ = \ \Psi''(0) \ = 0.
\label{finalPhi2}
\end{equation}


\subsection{The force and the rotation velocity.}
\label{subsec:velocity}

Once $\Psi(R)$ is known we can calculate $\Psi_{\rm eff}(R)$ 
using Eq. (\ref{psiphi}).
Equivalently, once $g(r)$ is known, $g_{\rm eff}(r)$ can be 
obtained through Eq.
(\ref{ggef}). Using these two equations together with Eq. 
(\ref{finalPhi}), 
and after some straightforward calculations, we can find a direct
relation between $g_{\rm eff}$ and $\Psi_{\rm eff}$:

\begin{equation}
	g_{\rm eff}(x)= 
		\Psi_{\rm eff}(x)-\frac{2}{\alpha^2}\Psi_{\rm eff}''(x)
		 + \frac{1}{\alpha^4}\Psi_{\rm eff}^{iv}(x) 
		 + \frac{4}{\alpha^2 x}\Psi_{\rm eff}'(x) 
		- \frac{4}{\alpha^4 x^4}[2x\Psi_{\rm eff}'(x)
		-2x^2\Psi_{\rm eff}''(x)+x^3\Psi_{\rm eff}'''(x)] 
\label{geffpsieff}
\end{equation}

From the behaviour of $\Psi$ at the origin (Eq.
(\ref{finalPhi2}), and Eq. (\ref{psiphi})), it is easy to see
that, at the origin, $\Psi_{\rm eff}$ will satisfy:

\begin{equation}
      \Psi_{\rm eff}(0) \ = \ \Psi'_{\rm eff}(0) \ = \ 
\Psi''_{\rm eff}(0) \ = \ 0.
\label{psieffcond}
\end{equation}

\section{Discusion and conclusions}
We have found the {\it exact} solution to the problem of 
inverting the integral
 relation between the elemental law of gravity and the gravitational
field generated by a spherical mass distribution. We have 
arrived at a direct and 
simple expression which can be useful in order to infer 
the law of 
gravity that could explain the large scale gravitational 
behaviour if we had 
good data on rotation curves of spherical galaxies with 
an exponential density 
profile. 
However, most of the observational data available on rotation
curves are for
 spiral galaxies. 
Despite the fact, shown in Appendix \ref{app:diskvsesf}, that this
formalism is a good approximation to study this class of galaxies
 if $g_{\rm eff}$ is a growing function of $r$, we believe that it is
better to study the case of a thin disk distribution that 
is a much better
approximation to the real morphology of spiral galaxies. 
In two forthcoming
publications   
we will give a similar solution for a thin-disk distribution 
(Rodrigo-Blanco \cite{Rodrigo96}) and we will
apply it to real galaxies (Rodrigo-Blanco \& P\'erez-Mercader 
\cite{RodrigoMercader96}).

\appendix
\section{Appendix: Gravitational difference between a 
disk and a sphere}
\label{app:diskvsesf}

\begin{figure*}
\setbox0=\hbox{
\epsfxsize=11cm
\epsfbox{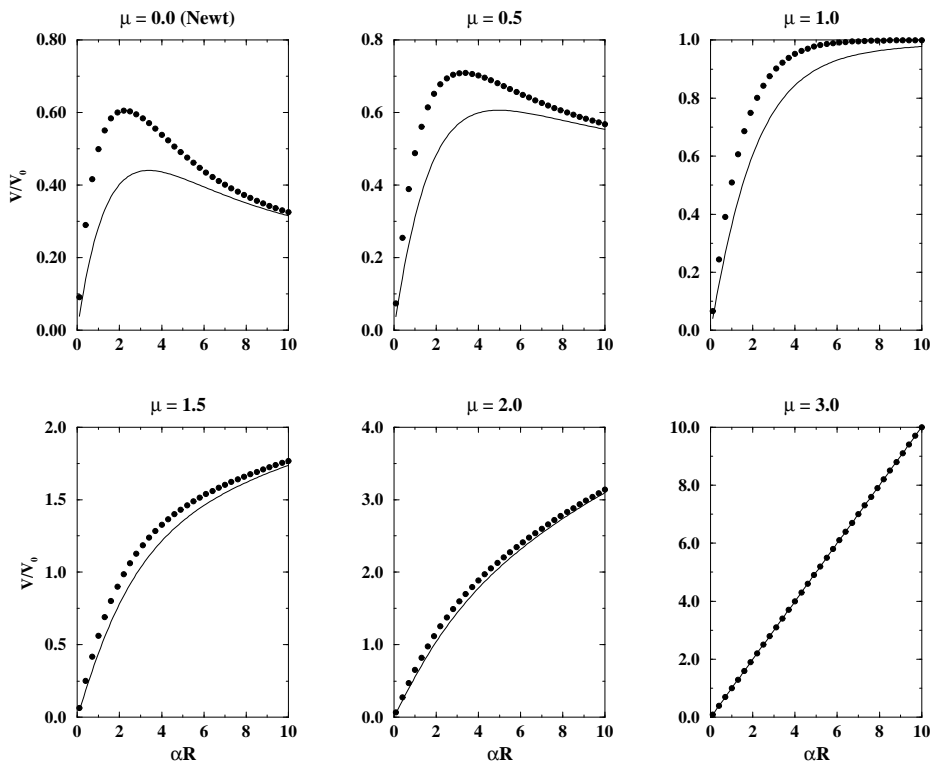}
}
\rotr 0

\caption{Rotation velocities for a sphere (solid line) and for a
disk (dots) when $g_{\rm eff}(r) \equiv
\left( \frac{r}{a}\right)^{\mu}$ for some values of $\mu$. In every
case, for the sake of clarity, the velocities are normalized by
dividing
by the appropiate constant 
$\frac{G_0 M_{\rm tot}\alpha}{(\alpha a)^{\mu}}$.}
\label{discesf}
\end{figure*}

At first sight, we would think that we cannot apply the results
of the previous sections to spiral galaxies,
because they have been obtained assuming spherical symmetry, and spiral
galaxies are disk-shaped, not spherical. In spite of this, it seems
logical that the difference between the gravitational field generated
by a sphere and the one generated by a disk becomes ne\-gli\-gi\-ble at
very large distances (here ``large distances" means large
compared with some typical length scale for the distribution; 
it could be, for
instance, $\alpha^{-1}$, i.e, the exponential length scale of the mass
distribution). Furthermore, when we consider the effect of an
increasing $g_{\rm eff}(r)$, it is evident that the meaning of 
what is a ``large
distance" changes as we change the functional form of 
$g_{\rm eff}(r)$. That is,
the faster $g_{\rm eff}(r)$ grows as a function of $r$,
 the smaller the distance
 at which a sphere is indistinguishable from a disk becomes, 
when considered
from the gravitational point of view.

In order to obtain a more quantitative idea of the gravitational 
difference 
between a sphere and a
disk, when $g_{\rm eff}$ is a growing function of its argument, 
we have chosen a 
parametric family of $g_{\rm eff}$ 's given by:

\begin{equation}
	g_{\rm eff,\mu}(r) \equiv \left(\frac{r}{a}\right)^{\mu}
\label{geffmu}
\end{equation}
where $\mu$ parameterizes how fast $g_{\rm eff}$ grows. 

Then we have calculated the rotation
velocity performing the three-dimensional integral numerically for
se\-ve\-ral values of $\mu$. We have
done it both for the case of spherical symmetry, and for the case 
of cylindrical symmetry. We have assumed for the disk a small
thickness of $h=\alpha^{-1}/6 $, and the rotation velocity is 
calculated in
the plane of the disk. 

In figure  (\ref{discesf}), we have plotted
both velocities for a disk as well as for a sphere 
for some values of $\mu$. In each case
we have normalized the solutions dividing by 
a convenient constant $V_0$ defined as:

\begin{equation}
V_0 \equiv \frac{G_0M_{\rm tot}\alpha}{(\alpha a)^{\mu}}
\label{defV0}
\end{equation}

It can be seen that both curves tend to merge as $\mu$
increases.

However, we would like to have a more
quantitative way of describing the difference between both curves as a
function of $\mu$. In order to do this, we define a quantity 
$\sigma^2_D$ 
(see below for its meaning) as follows:

\begin{equation}
	\sigma^2_D (\mu) \equiv \frac{1}{N} \sum_{\rm i=1}^{N} 
	\frac{(V_{\rm D,\mu}(r_i)-V_{\rm S,\mu}(r_i))^2}
	{V^2_{\rm D,\mu}(r_i)}
\label{sigmaDdef}
\end{equation}

Here the subscripts $D$ and $S$ stand for `disk' and `sphere',
res\-pec\-ti\-ve\-ly. We sum over $r_i$, which are the points where the
integrals are calculated. The total number of points for each 
value of $\mu$ is $N=100$.

\begin{figure}
\setbox0=\hbox{
\epsfxsize=7cm
\epsfbox{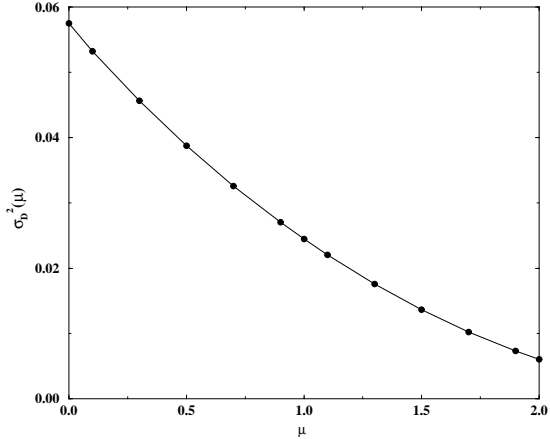}
}
\rotr 0

\caption{Mean square error ($\sigma^2_{\rm D}$) in the velocity 
as a function of $\mu$
when a sphere is considered instead of a disk,
 and using $g_{eff}(r) \equiv \left(\frac{r}{a}\right)^{\mu}$. 
The exact
meaning of $\sigma^2_{\rm D}$ is explained in the text.}
\label{sigmaD}
\end{figure}

Because of the way it is defined, $\sigma^2_D(\mu)$ is a 
measure of the mean square
 error that we make in the rotation velocity if we 
consider a sphere instead of the real disk, for each value of
 $\mu$. We thus aim for a value  as small as posible for $\sigma^2_D$.
 In Figure (\ref{sigmaD}), we plot the value
of $\sigma^2_D$ versus $\mu$. Once again, it can be seen that the
larger the value of $\mu$ is, the smaller the difference.

Considering what these plots mean, we see that
our mathematical results can be used in the case of spiral galaxies,
 with a good level of confidence, provided $g_{\rm eff}(r)$ grows
fast enough with $r$. Actually, the most popular corrections to 
$g_{\rm eff}=1$ that
can be found in the literature are $g_{\rm eff}\propto r$, (see 
Tohline \cite{Tohline83} and Khun \& Kruglyak \cite{KK87}), 
and $g_{\rm eff}\propto r^2$, (see Mannheim \& Kazanas 
\cite{Mannheim89}). 
It can be seen in the figures
that the approximation is quite good in both cases.
Although the
results obtained in  this way are not \mbox{exact}, 
they give a qualitatively correct picture. However, it 
will be better to use the solution to the problem in the case of disk 
symmetry itself, as we will do in a forthcoming publication
(Rodrigo-Blanco \cite{Rodrigo96}).

\section{Appendix: Mathematical identities}
\label{app:math}
  \begin{enumerate}
  \item{ Fourier sine transform}
	\begin{equation}
	 g(\sqrt{r^2+R^2-2rR\cos{\theta}}) \equiv 
	\frac{2}{\pi}
	  \int_0^{\infty}\hat{g}_s(p) \sin(p\sqrt{r^2+R^2-2rR\cos
	  {\theta}}) dp 
	\label{fourier}
	\end{equation}

	\begin{equation}
	  \hat{g}_s(p) = \int_0^{\infty} g(t) \sin{pt} dt 
	\label{fourierinv}
	\end{equation}

  \item{ Addition theorem for Bessel functions} (see Gradshteyn
\cite{Gradshteyn80}).

	\begin{equation}
	    	\frac{Z_{\nu}(m\omega)}{\omega^{\nu}}= 
	\frac{2^{\nu}}
    	{m^{\nu}}\Gamma(\nu)\sum_{\rm k=0}^{\infty}(\nu +k)
	       \frac{J_{\rm \nu+k}(m\rho)}{\rho^{\nu}} 
\frac{J_{\rm \nu+k}(mr)}
	   	{r^{\nu}}C_k^{\nu}(\cos{\theta})  
	\label{addth}
	\end{equation}
	where:
	\begin{equation}
	   	  \left\{\begin{array}{l}
		\omega \equiv \sqrt{r^2+\rho^2-2r\rho\cos{\theta}} \\
		\rho < r \\
		C_k^{\nu} \equiv \mbox{Gegenbauer Polynomials.}
		\end{array}\right. 
	\label{definitions}
	\end{equation}

\vspace{4mm}

\noindent	Using Eq. (\ref{addth}), together with:
	\begin{equation}
	\sin (mz) = \sqrt{\frac{\pi m z}{2}} J_{\rm 1/2}(mz)
	\label{sinbessel}
	\end{equation}

\vspace{4mm}

\noindent	we get:

	\begin{equation}
	\frac{\sin{p\sqrt{r^2+R^2-2rR\cos{\theta}})}}
	 {\sqrt{r^2+R^2-2rR\cos{\theta}}} = 
	\frac{\pi}{2\sqrt{Rr}}\sum_{\rm k=0}^{\infty}
	 (2k+1) J_{\rm k+\frac{1}{2}}(pr) 
	J_{\rm k+\frac{1}{2}}(pR) P_k(\cos{\theta})
	\label{addth2}
	\end{equation}

  \item{ Orthogonality of Legendre Polynomials.}
	\begin{equation}
	\int_0^{\pi} d\theta P_k(\cos{\theta}) \sin{\theta} =
	2\delta_{\rm k,0}
	\label{ortog} 
	\end{equation}

  \end{enumerate}

\acknowledgements{I want to thank Juan P\`erez-Mercader for his 
help and guidance in this work.}

\end{document}